\documentclass[conference]{IEEEtran}

\usepackage[papersize={8.5in,11in}, left=0.75in, right=0.75in, top=0.8in, bottom=0.8in]{geometry}

\ifCLASSINFOpdf

\else

\fi

\usepackage[utf8]{inputenc}

\usepackage{lettrine}
\usepackage{amsmath}
\usepackage{multirow}
\usepackage{graphicx}
\usepackage{epstopdf}
\usepackage{color}
\usepackage{placeins}
\usepackage{float}
\usepackage{tabularx,colortbl}

\usepackage{hyperref}

\begin{document}

%

\newcommand{\MYfooter}{\smash{
\hfil\parbox[t][\height][t]{\textwidth}{\centering
\thepage}\hfil\hbox{}}}

\makeatletter

%
%


\makeatother
\pagestyle{headings}
\addtolength{\footskip}{0\baselineskip}
\addtolength{\textheight}{-1\baselineskip}

%
\title{Hardware-Software Codesign for Software Defined Radio: IEEE 802.11p receiver case study}

\author{
\IEEEauthorblockN{
R. Zitouni$^{\ddagger}$, H. Bouaroua and B. Senouci
}
\IEEEauthorblockA{
  ECE Paris-LACSC Laboratory \\
  VEDECOM institute$^{\ddagger}$\\
37 Quai de Grenelle, 75015, Paris, France\\
Email:  rafik.zitouni@ece.fr$^{\ddagger}$ }
}

\maketitle


\begin{abstract}

Software Defined Radio (SDR) platforms are useful tools to design new wireless technologies or to improve specifications of existing ones. The IEEE 802.11p is the de-facto standard for Wireless Vehicular Ad-hoc NETworks (VANETs). It has been implemented on GNU Radio SDR \cite{Bloessl2013}, which experiences frames decoding/encoding latency. In this paper an FPGA based SDR is proposed as a solution to accelerate frame decoding of IEEE 802.11p waveforms. We share our experience in designing and validating an SDR on the top of real FPGA based embedded hardware architecture. On the top of an FPGA Zynq platform, we port an SDR GNU Radio following hardware/software (HW/SW) codesign approach. A real time profiling of the SDR system shows that the OFDM equalizer with FFT software function is the most time consuming task of the whole SDR receiver's functions (34.74\%). We suggest hardware acceleration of the FFT processing function using the free logic gates of the FPGA Zynq as a Hardware (HW) accelerator. The HW OFDM Equalizer with FFT sub-function could accelerate the SDR chain processing in comparison with a pure software implementation. Our FPGA based SDR design is a key step toward proposing an embedded SDR. It would be an alternative for not only the IEEE 802.11p standard but for any OFDM based SDR.

\end{abstract}

\begin{IEEEkeywords} Software Defined Radio (SDR), HW/SW codesign, FPGA-based Hardware Acceleration, Embedded Operating Systems, IEEE 802.11p standard, OFDM
\end{IEEEkeywords}



%
\IEEEpeerreviewmaketitle

\section{Introduction}
\lettrine[findent=2pt]{S}{ }OFTWARE Defined Radio platforms are software and hardware toolkits that allow designers to build reconfigurable wireless devices. SDR concept \cite{Mitola1993} has been introduced in order to migrate hardware (Hw) components, such as mixers, filters, amplifiers, modulators/demodulators, detectors, etc. from Hardware (Hw) to software (Sw) on a personal computer or an embedded architecture. GNU Radio \cite{Blossom2004} is a Swiss Army knife, an open-source toolkit allowing wireless communication designers to build software radios that run on General Purpose Processor (GPP). This toolkit can be used with an external RF hardware to create a Software Defined Radio (SDR), \textit{e.g.} Universal Software Radio Peripheral (USRP) \cite{Ettus} and BladeRF \cite{bladerf}, or without hardware in a simulated environment. GNU Radio provides signal processing blocks for modulation, demodulation, filtering and various data processing operations. New blocks can be added to experiment and to create new wireless point-to-point communication techniques. These blocks are developed in python or C++ programming languages and glued building flow graphs or transmitter/receiver chains.   

One possible use case of an SDR is by considering the vehicles as radio nodes of Wireless Vehicular Ad-hoc NETworks (VANETs). Radio VANETs allow cars to exchange information about their perception of the surrounding environment and their present position, speed, direction, etc, so that cars can cooperate in order to construct a Local Dynamic Map (LDM). In this way, potentially dangerous situations can be avoided. For example, recent crashes of autonomous cars would have been avoided if the involved vehicles had exchanged the appropriate information. The IEEE 802.11p protocol (and its European version called ITS-G5) \cite{ieee} has been specified for VANETs. It defines the MAC and Physical layers of vehicular networks operating in the 5.9 GHz band based on Orthogonal Freqeuncy Division Multiplexing (OFDM) method. An implementation of the IEEE 802.11p has been performed on GNU Radio \footnote{https://github.com/bastibl/gr-ieee802-11.git}  \cite{Bloessl2013}. It has been tested and found interoperable with hardware receiver and transmitter. 

USRP boards from Ettus Research \cite{Ettus} is a good candidate to perform a Front End (FE) radio offering high baseband sampling. Although the USRP holds a Field Programmable Gate Array (FPGA), it needs a host computer to perform baseband processing using high level programming (\textit{i.e.} C++ and Python). Since GNU Radio blocks exhibit parallelism, it is possible to map them totally or partially on an FPGA co-processor architecture. The benefits behind this mapping is mainly to improve the processing time of some blocks exploiting the parallel processing power of FPGAs. In addition, SDR designer can rid of host computer using only USRP boards as standalone SDR platform. RFNoc - RF Network-on-Chip \cite{Braun16} is a new open source framework similar to GNU Radio, but for developing applications which run on the FPGA. The USRP E310 is the embedded stand-alone SDR platform designed by Ettus Research \cite{EttusE310}. It contains a baseband processor, Xilinx Zynq ZC702 SoC (System on Chip) combined with a dual core ARM CPU, both into a single chip. The cost of such a board is expensive for hobbyists and users of free open source solutions. ZC702 \cite{ZC702} board contains a similar FPGA as USRP E310 but without a front end radio. It is affordable and it costs not more than 900 \$. Our objective is to deploy GNU Radio platform on this board. Then we would test and analyze the performances of GNU Radio flow graphs chains.

Many works have been published to make use of GNU Radio on FPGA based embedded plateform. Obviously, the subject is not purely scientific but its technical benefit allows researchers to have specific information in order to prototype efficiently proof of concepts. In \cite{J.Pendlum2013}, J. Pendlum has presented a first thought and implementation of GNU Radio block to support FPGA co-processing with Xilinx’s Zynq SoC. Finite Impulse Response (FIR) filter has been accelerated. Obtained results showed that the FPGA FIR filter processes samples five times quicker than FIR filter for the same number of taps. However, software and hardware setups are out of date or obsolete as announced in \cite{ZynqGR}. Moreover, the authors don’t show the consumed memory and processing time of each block, despite the existing performance counter tools such as \textit{ControlPort} \cite{Rondeau2013}. This latter is software function allowing GNU Radio designers to count processor and memory consumption of flow graphs. It creates an integrated remote procedure via Ethernet network. This procedure doesn't introduce any extra processing resource usage. One of our objective is to use \textit{ControlPort} tool within GNU Radio on OpenEmbedded Linux distribution \cite{OpenembeddedCore} running on ZC702 board. It would allow us to evaluate and compare the benefit of the implemented FIR filter \cite{J.Pendlum2013}. After that, we will extend ZC702 board usage by migrating other blocks. In our application, the block to migrate is a Fast Fourier Transform (FFT) block, which is a step in the OFDM demodulator of an IEEE 802.11p receiver. The result should accelerate the frame decoding. Note that this acceleration should be useful for not only our application, but for transceivers using FFT and Inverse FFT (IFFT).


The remainder of this paper is organized as follows. In Section II, we give a background about our tools and method. Then, we will present our case study and experiment results measuring the processor utilization. In the section III, we present an example of hardware acceleration \textit{i.e.}, FIR acceleration, by reporting performance analysis using GNU Radio tools. Our FFT hardware acceleration will be detailed and discussed in Section IV and finally, we give some conclusions and future directions in Section V.

\section{Software/Hardware tools}
This section gives a background about our Software Defined Radio, embedded systems and hardware acceleration method. GNU Radio, USRP B210, Zynq SoC and OpenEmbedded are our hardware and software tools.

\subsection{GNU Radio: Flow graphs}
GNU Radio is an open-source project toolkit for building software radios that run on GPP \cite{Blossom2004}. It is based on Mitola' s \cite{Mitola1993} concept by shifting ADC and DAC, \textit{i.e.} Analog/digital conversions, as close as possible to the antenna. The toolkit provides signal processing blocks for modulation, demodulation, filtering and various data and signal processing operations. New blocks could be easily programmed and integrated to the toolkit. Software radio programs are the result of block’s interconnection forming flow graphs. Each block can be developed in Python or C++ programming language. The flow graphs could simulate wireless communication chains or inter-operate with hardware transmitters/receivers in real-world scenario.

The benefit of GPP based flow graphs is a relative facility in high level programming languages. However, the latency is the most undesirable drawback if processing chains are with separated blocks. This latency comes from the scheduling and buffering of individual processing blocks (see Section. \ref{Sec:CaseStudy}). Typically, block processing is a function that reads a data stream from memory, processes it, and stores the output back to memory. One idea would be to migrate blocks from GPP to an FPGA when designers don’t expect high level of reconfiguration. By contrast, the FPGA is powerful and available when we use hybrid architecture, \textit{i.e.} GPP with a hardware programmability of an FPGA, such as Zynq boards. 

\textit{ControlPort} \cite{Rondeau2013} is software tool based on Linux performance counters which allows us to measure the performances of GNU Radio flow graphs and in particular the IEEE 802.11p receiver. The concept is useful in order to debug and to measure the CPU and memory usage of each flow graph's block. In fact, samples are buffered between each software processing block. \textit{ControlPort} should allow us to identify the complicated algorithms consuming most of the processing time. This delay has an impact on both, the physical layer of protocol stacks and also on a phase synchronization loop. Indeed, flow graphs control Hardware RF front end for real wireless communications. Universal Software Radio Peripheral (USRP) \cite{Ettus}, RealTeck SDR and HackRF are candidate devices. Ettus (USRPs) remain the promising devices with high performances, such as B210, N210 and E313. In our work, we choose a USRP B210 covering a large frequency bandwidth from 70 MHz to 6 GHz. 

\begin{figure*}
\centering
\includegraphics[width=6.8in]{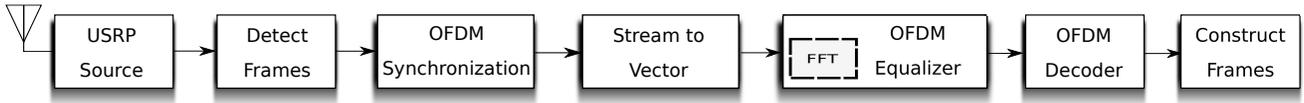}
\caption{Simplified GNU Radio flow graph of IEEE 802.11p receiver}
\label{fig:FGReceiver}
\end{figure*}


%




\subsection{Zynq SoC}
The Zynq is a new Xilinx programmable platform for embedded SoC and HW/SW systems  \cite{ZC702} \cite{Crockett2014}. It is a family of FPGA with different hardware characteristics but having the same basic architecture. As shown in Fig. \ref{fig:HwArchitecture}, this chip comprises two parts: the Processing System (PS), and the Programmable Logic (PL). The PS is two hardware processor ARM Cortex-A9. The components of the ARM processor are a set of associated processing resources forming an Application Processing Unit (APU), and further peripheral interfaces, cache memory, memory interfaces, interconnect, and clock generation circuitry. The PL is based on the Artix-7 and Kintex-7 FPGA and predominantly composed of general purpose FPGA logic fabric with slices and Configurable Logic Blocks (CLBs), and Input/Output Blocks (IOBs) for interfacing. The PS and PL can be used independently or together. In fact, the power circuitry is configured with separate domains for each, enabling either the PS or PL to be powered down if not in use. The most compelling model for Zynq is when both parts are used in conjunction. The connection between them is achieved by communication bus named AXI (Advanced eXtensible Interface). An efficient implementation of GNU Radio flow graphs would be to migrate the signal processing blocks (FIR filter, FFT or IFFT) into the PL part instead of running them on the GPP. The processing time with an FPGA would be greatly reduced.

With a such SoC architecture, GNU Radio's flow graphs can process In/output digital streams through a number of blocks located in both FPGA and ARM processors. Thus, we need to customize an Embedded Linux operating system with its hardware drivers ensuring stream exchange between blocks of FPGA and GPP.   


\begin{figure}
\centering
\includegraphics[width=3.4in]{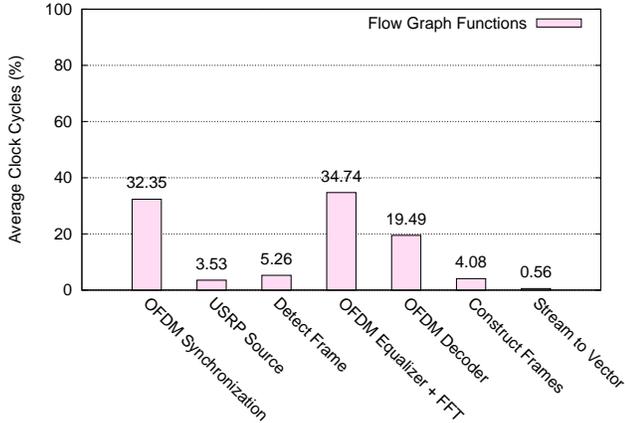}
\caption{Processor utilization by function of receiver flow graph}
\label{fig:FGClockCycles}
\end{figure}

\subsection{OpenEmbedded and Bitbake}
The OpenEmbedded \cite{OpenembeddedCore} is an open-source project allowing software designers to customize a framework for embedded Linux. This framework contains tools, metadata, recipes, etc, to build and configure a complete Linux system. To run a given configuration we need the \textit{Bitbake} tool \cite{Bitbake}, which is a shell script with specific commands to execute the different tasks of building and compiling the Linux image.  Linux distribution can be customized for different embedded processor architectures, \textit{e.g.} ARM, MIPS, x86 .etc, and for needed software tools and libraries \textit{e.g.} gcc, gdb, Qt, etc (see Section \ref{Sec:Bitbake}).
Our objective is to build a customized OpenEmbedded Linux with GNU Radio running on Zynq SoC. This distribution would support GNU Radio toolkit with \textit{ControlPort} tool.


\begin{figure*}
\centering
\includegraphics[width=5in]{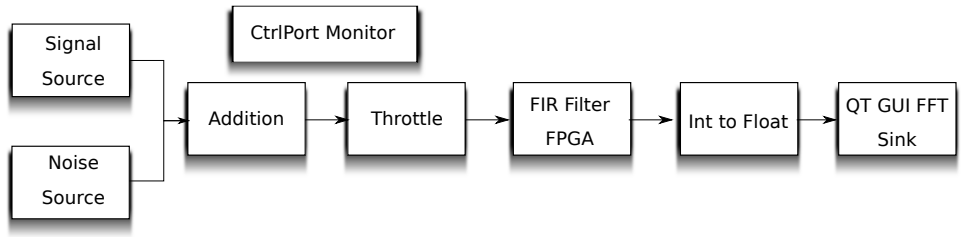}
\caption{Flow graph with an FPGA FIR filter}
\label{fig:flowgraph}
\end{figure*}

\section{Software Defined Radio for IEEE 802.11p}
\label{Sec:CaseStudy}
SDR platforms or transceivers experience latency to meet the requirements of network protocols. From the MAC to the application layer of the open system interconnection (OSI) network stack, SDR devices of networks spend more time to encode or to decode data frames. In the IEEE 802.11p standard \cite{ieee}, the data broadcast is one possible operation. In fact, the vehicles cooperate by exchanging CAM messages (Cooperative Awareness Messages) for safety applications. In this case, the network has less timing constraints. The vehicles doesn't need Acknowledgment packets or handshake protocol. Thus, an SDR implementation would be an interesting solution when we need a maximum flexibility at physical layer. But our SDR has to considerate the unicast communication in particular cases, \textit{i.e.} send information from source  to a target vehicle. Therefore, vehicles need to use ACK (Acknowledgement) messages and CSMA/CA protocol to avoid hidden vehicles problem. Consequeuntly, the time to encode/decode frames should be minimized as possible in order to avoid performance leakage caused by head of line blocking phenomenon \cite{Klingler2015}. 

%
%
%
 
GNU Radio implementation of 802.11p standard is available as an open source project \cite{Fuxjager}\cite{Bloessl2013}. It implements the standard specifications as a transmitter and receiver flow graphs. The receiver flow graph defines software processing steps from baseband waveform reception to a data frame decoding.  Fig. \ref{fig:FGReceiver} shows an abstraction of the detailed version present in \cite{Bloessl2013}.

The OFDM is the scheme defined at the physical layer. The receiver starts by detecting the preamble sequence of data frames. After that, it synchronizes frequency and ensures symbol alignment applying frequency offset correction. To accelerate the software processing, we gather complex items in vectors to prepare equalization. It is possible to have an SIMD (Single Instruction Multiple Data) processing  of these vectors. Thereafter, the OFDM equalizer performs the phase offset correction and channel estimation. A phase offset is generated car the sampling times of the sender and the receiver are not synchronized and the symbol alignment is not perfect. An FFT operation is applied on the received digital stream. The two last steps are frames decoding and packet reconstruction, which allowed us to calculate packet error rate. Thus, our first objective is to measure the CPU and the memory consumption of the flow graph's blocks. Then, we would be able to identify the greedy blocks or most consumer of processing resources. 

\subsection{Experiments}
We have performed IEEE 802.11 frames reception with USRP B210 as RF front end and the GNU Radio flow graph running on a host computer (or GPP processor). We were be able to decode control, management and data packets of IEEE 802.11g. In addition, the flow graph has been found interoperable with autotalks transceivers \cite{AutoTalks} decoding CAM packets of IEEE 802.11p. Fig. \ref{fig:FGClockCycles} shows the percentage of  processor utilisation by each function of the receiver flow graph. Through \textit{gr-ctrlport-monitor} tool, we have calculated the average clock cyles of each flow graph's block. We found that the OFDM Equalizer uses higher average clock cyles than all other blocks. The transition step from time to frequency domain is done by an FFT block. Since FFT is a step that prepares samples to equalizer function, we gathered two blocks in OFDM equalizer function. This function introduces phase offset correction and channel estimation. The megration of FFT and OFDM equalizer function on an FPGA shall allows receiver to accelerate the frame decoding by 34.74\%.

Our method to migrate an FFT block is based on a design of an existing hardware acceleration example. We choose to analyze this example by doing a reverse engineering of FIR filter implementation on Zynq board  \cite{J.Pendlum2013}, which is presented in Section \ref{Sec:ReVFIR}. Since numerous software/hardware tools will be used, the next section gives a brief overview of these tools.

\section{FPGA-based FIR filter Hardware Acceleration}
\label{Sec:ReVFIR}
The hardware acceleration example has been obtained from \cite{J.Pendlum2013}. It is a GNU Radio FIR filter implemented on the Zynq SoC. The author has reused and adapted an Intellectual Propriety (IP) Core provided by \textit{Xilinx}. The available project source code and design have been found outdated and needed reworking. During our reverse engineering of this implementation, we fixed problems related to its setup by debugging the shared source code in \footnote{https://wiki.gnuradio.org/index.php/Zynq}. The reverse engineering of this example allowed us to master the process of design of FPGA-based hardware acceleration for an FFT block.

\begin{figure}
\centering
\includegraphics[width=3.4in]{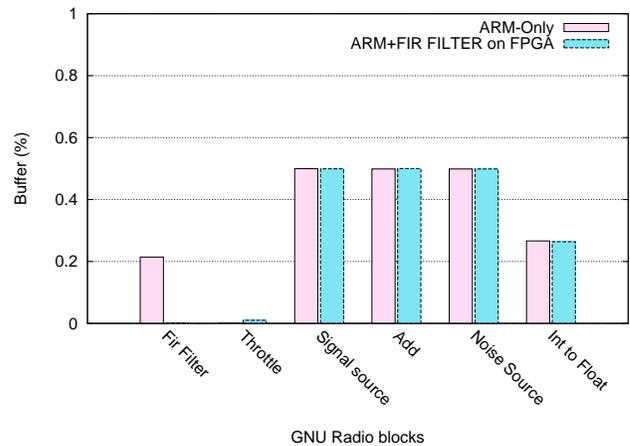}
\caption{Memory buffer occupation of the flow graph's blocks}
\label{fig:buffer}
\end{figure}

\subsection{FPGA FIR filter}
\label{sec:FIRFilter}
The flow graph depicted in Fig. \ref{fig:flowgraph} gives an example of FPGA FIR filter utilization. It is a loop back signal processing function running on combined ARM and FPGA architecture. Its result is the application of the low pass FIR filter and then the plot of its FFT. The first block generates cosine waveform with a given frequency, sample rate and signal amplitude. A Gaussian noise is also generated and after that added to the signal. The obtained sample stream enters through a \textit{Throttle} block in order to limit frequency speed of the CPU. Then, the obtained signal enters through a low pass FIR filter implemented on a Zynq FPGA. The output of hardware processing is caught by \textit{Int to Float} block running on ARM processor. The sink block plot an FFT of the processed stream. 


Obviously, this flow graph is expected to run on OpenEmbedded Linux Distribution. In fact, errors have been encountered when we built the existing project. They were related to the Linux Kernel and drivers. By debugging the existing project, we updated meta-files and we built a user-peripheral driver which responds to system calls. This problem had been resolved with ZC270 \cite{ZC702}. The Zynq's Accelerator Coherency Port (ACP) has been used by developing a kernel device driver and  Hardware Description Language (VHDL) interface. The FPGA block exchanges data with ARM CPU via the AXI Datamover block (see Fig. \ref{fig:HwArchitecture}). The authors implemented a kernel module (or driver) in order to control FPGA FIR filter through GNU Radio running on ARM CPU. The input and the output of this block are samples with integer data type. We have only Filter Taps as a user accessible parameter via Graphical User Interface (GUI) block.

The \textit{ControlPort} block was connected to the flow graph and allowed us to calculate resources utilization of each block. Fig. \ref{fig:buffer} show the memory buffer and CPU rate occupancy, respectively. The obtained results compare the performances of the same flow graph running only on ARM processor and on Zynq SoC. With a FIR filter on FPGA, the processing has been accelerated, and we saved 21\% of memory utilization. Obviously, the ARM processor has not been used and all ARM processor was shared by the remaining blocks. By contrast, only on ARM processor, all the blocks share the GPP architecture resources.

\section{System design}

This section presents our hardware-software codesign architecture. We will start from the configuration of our OpenEmbedded image generation to the calculation of FFT's resources utilization. 

\subsection{Bitbake Process}
\label{Sec:Bitbake}

\begin{figure}
\centering
\includegraphics[width=2.5in]{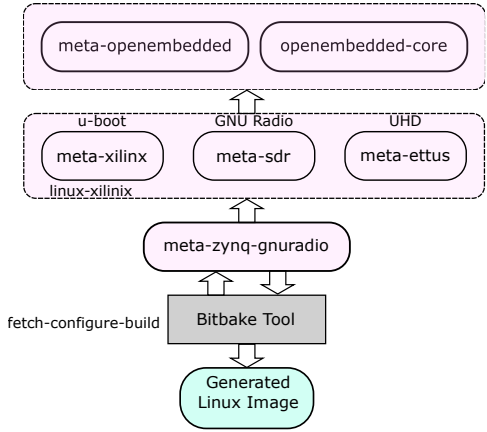}
\caption{Required meta-layers for Linux OpenEmbedded image}
\label{fig:BitbakeMeta}
\end{figure}

We present our system design from OpenEmbedded Linux image configuration to Hardware acceleration. The OpenEmbedded Linux image has been configured and generated via Bitbake tool. The latter comes in form of shell script with specific commands to execute the different tasks of building and compiling. The enters of the Bitbake are the meta-layers, for example, GNU Radio is a meta layer named \textit{meta-sdr} \footnote{https://github.com/balister/meta-oe.git}. Fig. \ref{fig:BitbakeMeta} shows required meta-layers included by Bitbake to describe how to setup the Gnu Radio on our Linux distribution. The \textit{meta-ettus} layer is used also to include the USRP Hardware Driver (UHD) which allows the GNU Radio to recognize boards connected to the Linux Embedded based platforms, in our case, it is ZC702. We make shared a repository of our modifications and indications to rebuild an OpenEmbedded linux image \footnote{https://github.com/zitouni/ZynqGNURadio.git}. You will check the meta sources \textit{meta-ettus}, {meta-sdr} and {meta-xilinx}. 

One other important meta-layer for building a specific Linux OS for the Zynq SoC is the \textit{meta-xilinx}. This layer contains the recipes for u-Boot \textit{Boot-loader}, the board support package (BSP) as well as the Linux-Xilinx distribution for this design. In fact, the \textit{meta-zynq-gnuradio} layer contains the essential recipes which indicate to the Bitbake tool the desired Linux distribution and the reference of Zynq SoC. Also this meta-layer defines the dependencies to the three essential meta-layers explained above, \textit{i.e.} \textit{meta-xilinx}, \textit{meta-sdr}, \textit{meta-ettus}, in order to include the required files. Note that the user can modify this meta layer or use it as an example in order to create his own customized Linux distribution. After building the Linux, we will have a complete Linux running on the Zynq SoC to develop GNU Radio flow graph with USRP as Radio Frequency (RF) FE.

\subsection{Software/Hardware configuration}

\begin{figure}
\centering
\includegraphics[width=3.2in]{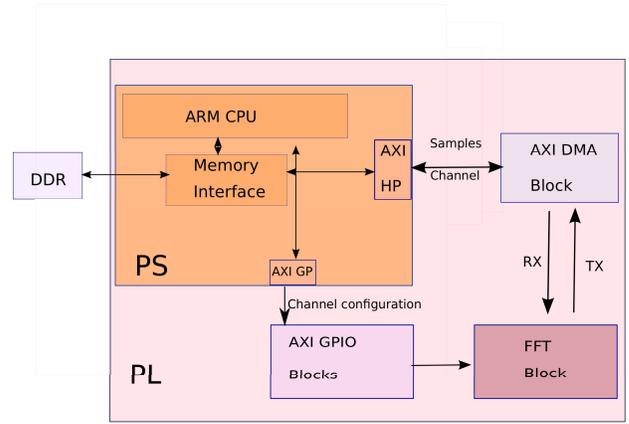}
\caption{Architecture ZC702 with the FFT IP as a hardware accelerator component for our SDR}
\label{fig:HwArchitecture}
\end{figure}

The FFT block is implemented on the Programmable Logic (PL), \textit{i.e.} \textit{FPGA fabric} side of the Zynq SoC, whereas the GNU Radio platform is supported on the Processing System (PS). Such architecture reduces the latency of a flow graph. Fig .\ref{fig:HwArchitecture} gives an overview of our Hardware/Software co-processing configuration. GNU radio platform is running on the ARM PS side as an application on the mounted OpenEmbedded image. The FFT block is an entity purely  in hardware. The GUI of hardware FFT block is implemented on GNU Radio. It exchanges samples with hardware implementation through an AXI bus. The latter ensure data samples exchanges between PS and PL domains. However, a device driver has been added to Linux kernel in order to establish data transfer between the two domains. Note that not only GNU Radio flow graph, \textit{i.e.} IEEE 802.11p receiver, can use the FFT block. Any other application can reuse this hardware FFT block. 

\subsection{Hardware setup}

Fig. \ref{fig:HwArchitecture} reports Hardware acceleration architecture, which has been designed on Vivado Xilinx suits software. The Hardware part contains the FFT IP core or Xilinx LogiCore IP, the AXI Direct Memory Access (DMA) and the AXI General Purpose Input Output (GPIO) IP blocks. The AXI DMA ensures data transfer between (FFT/IFFT resulted samples) and Processing system (ARM CPU) entity. The samples pass through the AXI High Performance (HP) bus. In addition, the AXI General Purpose (GP) allows us to control register access of the FFT block through the AXI GPIO IP block.

The AXI GP ports are useful for access memory mapped devices, since they have 32 bits wide and they are split on two slaves and two master ports. We enable one AXI GP port as master to read/write access and to control registers in the FFT block. This control register allows us to change a number of our FFT block such as inverse/forward (IFFT/FFT), number of points or FFT size, cyclic prefix, etc. However, the AXI HP ports are adapted to high throughput data transfer between ARM processor and FPGA fabric. They are similar to the AXI GP ports by splitting them into master/slave ports. The AXI HP ports can directly read and write to the CPU’s RAM via the Multiport DRAM memory interface. Each port is on 64-bits wide and capable of operating at 150 MHz. This results in a peak of 1.2 GB/sec transfer rate for both reads and writes.

If we compare our hardware setup to that proposed for the FIR filter (see Section  \ref{sec:FIRFilter}), the AXI ACP port has been replaced by the AXI HP port. With numerous GNU Radio blocks running on Linux (or GPP architecture), the processor cache is quickly full and DRAM space is required for data transfer. Thus, the performances of AXI HP ports could be powerful to those of Accelerator Coherency Port (ACP). The AXI DMA IP provides high bandwidth direct access memory between the AXI memory mapped IPs (ARM-CPU) and the (AXI-Stream IP) interfaces. Note that the benefits of the DMA block is to offload the CPU of moving data to/from the programmable logic IP, and let the DMA block to do the task ones is enabled through the configuration entry. Here the AXI DMA block add an address memory for the FFT block to be visible on the address range of the ARM CPU.

\subsection{Resources utilization}

On the HW side, the logical synthesis process is performed using Xilinx Synthesis Tool (XST) tool. It does take more than half an hour to complete synthesis the hardware architecture on the Zynq platform. Table. \ref{table:fpgautiliz} shows the synthesis logic results; where the component names refer to the HW modules . The other columns represent the logic size in terms of logic gates/Look-Up-Table, memory banks, DSP and FlipFlops used out of the 100\% capacity of the hardware resources of the Zynq ZC702. The hardware FIFO IP took less than 3\% of LUT slices. The remaining step is to connect our FFT hardware block to the intermediate blocks of the receiver's flow graph.

\begin{table} [!htb]
\caption{FPGA Resource utilization of our design on Zynq ZC702}
\centering
  \begin{tabular}{ | c | c | c | c | }
    \hline
    Resource  & Used & Available & Percent Used \%  \\ \hline
    Slice LUTs & 7995 & 53200& 15.03 \\ \hline 
    Slice Registers & 11867 & 106400 & 11.15  \\ \hline 
    DSPs & 18& 220  & 8.18 \\ \hline 

  \end{tabular}
\label{table:fpgautiliz}
\end{table}

\section{Conclusion}

In this paper, we reported our results implementing hardware acceleration for GNU Radio SDR. Our main contribution is the measured resources utilization of software functions decoding IEEE 802. 11p frames. We found that the OFDM equalizer was the highest consumer of processing time. Compared to other receiver's functions, the equalizer consumes 34.74\% of summed averaged clock cycles of all functions. We proposed to migrate FFT function from a GPP to an FPGA. The designed architecture has been detailed from the embedded system setup to the calculation of logic resource utilization on our Zynq SoC. Our migration is feasible since logic size utilization of our FFT hardware accelerator has been found lower than 50\% of all available resources. 

Our ongoing work is to fully integrate the accelerated FFT block in an IEEE 802.11p receiver and evaluate the SDR receiver performances. This step would be a daunting task with a lot of time should be spent setting up embedded operating system, clocking, configuring transports, etc.

\bibliographystyle{IEEEtran}
\bibliography{biblio}

\begin{thebibliography}{10}
\providecommand{\url}[1]{#1}
\csname url@samestyle\endcsname
\providecommand{\newblock}{\relax}
\providecommand{\bibinfo}[2]{#2}
\providecommand{\BIBentrySTDinterwordspacing}{\spaceskip=0pt\relax}
\providecommand{\BIBentryALTinterwordstretchfactor}{4}
\providecommand{\BIBentryALTinterwordspacing}{\spaceskip=\fontdimen2\font plus
\BIBentryALTinterwordstretchfactor\fontdimen3\font minus
  \fontdimen4\font\relax}
\providecommand{\BIBforeignlanguage}[2]{{%
\expandafter\ifx\csname l@#1\endcsname\relax
\typeout{** WARNING: IEEEtran.bst: No hyphenation pattern has been}%
\typeout{** loaded for the language `#1'. Using the pattern for}%
\typeout{** the default language instead.}%
\else
\language=\csname l@#1\endcsname
\fi
#2}}
\providecommand{\BIBdecl}{\relax}
\BIBdecl

\bibitem{Bloessl2013}
B.~Bloessl, M.~Segata, C.~Sommer, and F.~Dressler, ``{Towards an Open Source
  IEEE 802.11p stack: A full SDR-based transceiver in GNU Radio},'' in
  \emph{2013 IEEE Vehicular Networking Conference}.\hskip 1em plus 0.5em minus
  0.4em\relax IEEE, dec 2013, pp. 143--149.

\bibitem{Mitola1993}
J.~Mitola, ``{Software radios: Survey, critical evaluation and future
  directions},'' \emph{IEEE Aerospace and Electronic Systems Magazine}, vol.~8,
  no.~4, pp. 25--36, apr 1993.

\bibitem{Blossom2004}
E.~Blossom, ``{GNU radio: tools for exploring the radio frequency spectrum},''
  \emph{Linux J.}, vol. 2004, no. 122, pp. 4----, jun 2004.

\bibitem{Ettus}
\BIBentryALTinterwordspacing
``{Ettus Research}.'' [Online]. Available: \url{https://www.ettus.com/}
\BIBentrySTDinterwordspacing

\bibitem{bladerf}
\BIBentryALTinterwordspacing
``Bladerf.'' [Online]. Available: \url{http://nuand.com/}
\BIBentrySTDinterwordspacing

\bibitem{ieee}
I.~C. Society, \emph{Part 11: Wireless LAN Medium Access Control (MAC) and
  Physical Layer (PHY) Specifications}.\hskip 1em plus 0.5em minus 0.4em\relax
  IEEE Computer Society, 2012.

\bibitem{Braun16}
M.~Braun, J.~Pendlum, and M.~Ettus, ``Rfnoc: Rf network-on-chip,''
  \emph{Proceedings of the GNU Radio Conference}, vol.~1, no.~1, 2016.

\bibitem{EttusE310}
\BIBentryALTinterwordspacing
``{USRP E310 Embedded Software Defined Radio (SDR) - Ettus Research}.''
  [Online]. Available: \url{https://www.ettus.com/product/details/E310-KIT}
\BIBentrySTDinterwordspacing

\bibitem{ZC702}
\BIBentryALTinterwordspacing
``{Xilinx Zynq-7000 All Programmable SoC ZC702 Evaluation Kit}.'' [Online].
  Available:
  \url{https://www.xilinx.com/products/boards-and-kits/ek-z7-zc702-g.html}
\BIBentrySTDinterwordspacing

\bibitem{J.Pendlum2013}
{J. Pendlum}, ``{GSoC13: GnuRadio FPGA Co-processing with the Xilinx Zynq
  System-on-Chip},'' in \emph{Google Summer of Code}, 2013.

\bibitem{ZynqGR}
\BIBentryALTinterwordspacing
``{Zynq - GNU Radio - gnuradio.org}.'' [Online]. Available:
  \url{http://gnuradio.org/redmine/projects/gnuradio/wiki/Zynq}
\BIBentrySTDinterwordspacing

\bibitem{Rondeau2013}
T.~W. Rondeau, T.~O'Shea, and N.~Goergen, ``{Inspecting GNU radio applications
  with controlport and performance counters},'' in \emph{Proceedings of the
  second workshop on Software radio implementation forum - SRIF '13}.\hskip 1em
  plus 0.5em minus 0.4em\relax New York, New York, USA: ACM Press, 2013, p.~65.

\bibitem{OpenembeddedCore}
\BIBentryALTinterwordspacing
``{OpenembeddedCore-Yocto Project}.'' [Online]. Available:
  \url{https://www.yoctoproject.org/tools-resources/projects/openembeddedcore}
\BIBentrySTDinterwordspacing

\bibitem{Crockett2014}
L.~H. L.~H. Crockett, R.~A. Elliot, M.~A. Enderwitz, and R.~W.~E. engineer)
  Stewart, \emph{{The Zynq book : embedded processing with the ARM Cortex-A9 on
  the Xilinx Zynq-7000 all programmable SoC}}, 2014.

\bibitem{Bitbake}
\BIBentryALTinterwordspacing
``Bitbake.'' [Online]. Available:
  \url{https://www.yoctoproject.org/tools-resources/projects/bitbake}
\BIBentrySTDinterwordspacing

\bibitem{Klingler2015}
F.~Klingler, F.~Dressler, and C.~Sommer, ``{IEEE 802.11p unicast considered
  harmful},'' in \emph{2015 IEEE Vehicular Networking Conference (VNC)}.\hskip
  1em plus 0.5em minus 0.4em\relax IEEE, dec 2015, pp. 76--83.

\bibitem{Fuxjager}
\BIBentryALTinterwordspacing
P.~Fuxj{\"{a}}ger, P.~Fuxj{\"{a}}ger, A.~Costantini, D.~Valerio,
  P.~Castiglione, G.~Zacheo, T.~Zemen, and F.~Ricciato, ``{1 IEEE 802.11p
  Transmission Using GNURadio}.'' [Online]. Available:
  \url{http://citeseerx.ist.psu.edu/viewdoc/summary?doi=10.1.1.192.1722}
\BIBentrySTDinterwordspacing

\bibitem{AutoTalks}
\BIBentryALTinterwordspacing
``{Auto Talks}.'' [Online]. Available: \url{http://www.auto-talks.com/}
\BIBentrySTDinterwordspacing

\end{thebibliography}

\end{document}